\documentclass{iau} 
\usepackage{graphicx}
\usepackage{natbib}

\newcommand{\apj}{ApJ}
\newcommand{\aap}{A\&A}
\newcommand{\mnras}{MNRAS}
\newcommand{\nat}{Nature}
\newcommand{\apjl}{ApJL}
\newcommand{\apjs}{ApJS}
\newcommand{\aj}{AJ}

\newcommand{\araa}{Ann. Rev. Astron. Ast.}

\title[Invited review: Tidal effects on stellar activity] 
{Tidal effects on stellar activity}

\author[K.~Poppenhaeger]   
{K.~Poppenhaeger$^{1, 2}$}

\affiliation{$^1$Astrophysics Research Centre, Queen's University Belfast, BT7 1NN Belfast, United Kingdom \\ 
email: {\tt k.poppenhaeger@qub.ac.uk} \\[\affilskip]
$^2$Harvard-Smithsonian Center for Astrophysics, 60 Garden Street, Cambridge, 02138 MA, USA}

\pubyear{2017}
\volume{328}  
\jname{Living around active stars}
\editors{ }
\begin{document}

\maketitle

\begin{abstract}
The architecture of many exoplanetary systems is different from the solar system, with exoplanets being in close orbits around their host stars and having orbital periods of only a few days. We can expect interactions between the star and the exoplanet for such systems that are similar to the tidal interactions observed in close stellar binary systems. For the exoplanet, tidal interaction can lead to circularization of its orbit and the synchronization of its rotational and orbital period. For the host star, it has long been speculated if significant angular momentum transfer can take place between the planetary orbit and the stellar rotation. In the case of the Earth-Moon system, such tidal interaction has led to an increasing distance between Earth and Moon. For stars with Hot Jupiters, where the orbital period of the exoplanet is typically shorter than the stellar rotation period, one expects a decreasing semimajor axis for the planet and enhanced stellar rotation, leading to increased stellar activity. Also excess turbulence in the stellar convective zone due to rising and subsiding tidal bulges may change the magnetic activity we observe for the host star. I will review recent observational results on stellar activity and tidal interaction in the presence of close-in exoplanets, and discuss the effects of enhanced stellar activity on the exoplanets in such systems.

\end{abstract}

\firstsection 
\section{Stellar activity}

Stellar activity is a collective term for a variety of magnetic phenomena observed in cool stars, i.e.\ stars with outer convective envelopes (spectral types mid-F to mid-M) or stars that are fully convective (mid-M and later). Manifestations of magnetic activity include the presence of flares, coronal mass ejections, chromospheres and coronae, starspots, and faculae. 


All of these phenomena are ultimately driven by the stellar rotation through the magnetic dynamo \citep{Parker1955}. The differential rotation of a star, both latitudinally and radially, causes not only a long-term inversion of the global magnetic field polarity, but also the localized phenomena in the stellar atmosphere which make up the individual facets of magnetic activity. Understanding the evolution of stellar rotation, from the formation of stars through the gigayears of their lifetime, is therefore fundamental to our understanding of stellar magnetic activity.


One important factor of the rotational evolution of cool stars the the spin-down that occurs due to magnetic braking \citep{Schatzman1962}. This happens because cool stars shed an ionized stellar wind, which moves away from the star along the stellar magnetic field lines. Finally, it decouples from the magnetic field, and at this moment the angular momentum is carried out of the system by the stellar wind. This continuous loss of angular momentum causes a spin-down of the star over time, which can be studied observationally \citep{Barnes2003, Barnes2010, Meibom2015, vanSaders2016}. Typically, stars set out on the main sequence with short rotation periods of half a day to a few days, and spin down to long periods of ca.\ 30 days over a few gigayears in the case of the Sun, and much longer rotation periods of the order of 100 days for low-mass stars on the main-sequence at old ages \citep{Irwin2011}. Studying slow rotation of stars is observationally challenging, because the main observables of rotation (rotational broadening of spectral lines and photometric variability due to star spots on the stellar surface) provide only weak signatures in the slow-rotation regime.


As rotation slows down, the stellar magnetic activity decreases. While this qualitatively makes sense, since (differential) rotation is the motor for stellar activity, the physical details of this are not fully understood. For example, it is not clear how the rotational period of a star and the presence and duration of activity cycles (i.e.\ the 11-year activity cycle of the Sun) are related. Still, the overall effects of activity, such as coronal X-ray emission \citep{Guedel1997, Preibisch2005, Telleschi2005}, chromospheric line emission such as the Ca II H and K lines and the H alpha line \citep{Skumanich1972, Noyes1984, Mamajek2008, Reiners2012}, and photospheric variability \citep{Bastien2014, Stelzer2016} can be related to stellar rotation and also directly to stellar age. 

However, magnetic braking is not the only physical effect that influences stellar rotation and activity over time, which brings us to tidal interaction.

\section{Tidal interaction}

Whenever we have two astronomical objects in close proximity to each other, tides start to play a role. For our topic of interest, examples for relevant systems are: a close binary system consisting of two stars, a star with a massive planet in a close orbit, or a planet-moon system. Tides cause a deformation of the involved bodies due to the gravitational force acting on them, and due to their rotation around their common center of mass. There are three main observational effects of tides in such systems: alignment of the spin axes perpendicular to the orbital plane; synchronization, meaning that over time the rotational periods of the bodies and the orbital period become equal; and circularization, meaning that bodies in an eccentric orbit slowly lose their eccentricity and adopt a circular orbit \citep{Zahn2008, Mathis2009}.

\begin{figure}
\includegraphics[width=\textwidth]{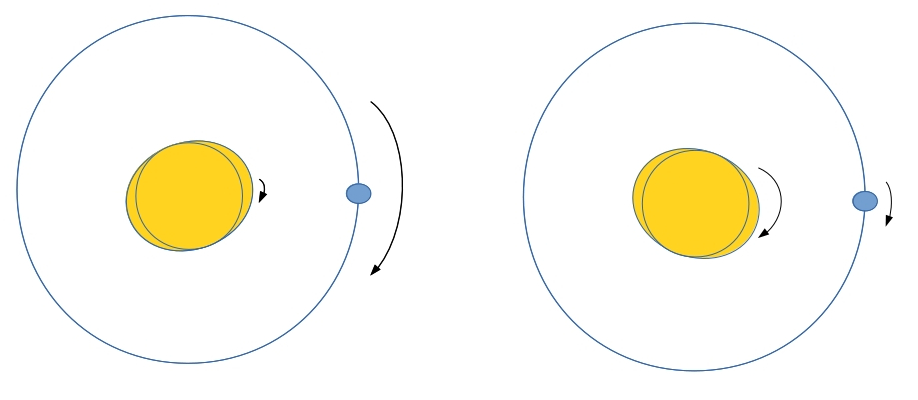}
\caption{Example of tidal interaction in a two-body system. In the example to the left, the rotation period of the central object (A) is longer than the orbital period of the smaller body (B), in the example to the right the rotation period of the central object is shorter. Typically, for star-planet systems with Hot Jupiters and relatively old host stars, the left example is representative of the orbital and rotational periods observed.}
\label{tidal}
\end{figure}

For our topic of stellar activity, the synchronization effect is the most relevant one. Therefore we take a more detailed look at how this effect plays out in different combinations of orbital and rotational periods. Let us assume a system of two bodies A and B in a close-in orbit, as depicted in Fig.\ \ref{tidal}. Both objects get deformed by tides, and here we focus on what happens to the central body (A) due to the tides. 

Assume that A has a longer rotational period than the orbital period of B. B raises a tidal bulge on A, and because B moves faster on its orbit than A rotates (in terms of angular velocity), the tidal bulge on A will lag behind (see Fig.\ \ref{tidal} left side). The gravitational pull of B on this bulge will therefore induce a tidal torque, and pull A into a somewhat faster rotation. Angular momentum is transferred from the orbit of B to the spin of A. Since the total angular momentum of the system is conserved, the semi-major axis of B decreases as the angular momentum of its orbit decreases, meaning B spirals slowly closer to A. 

In the opposite configuration, where A has a shorter rotational period than the orbit of B, angular momentum is transferred into the other direction. The tidal bulge on A runs ahead, and gets ``pulled back'' by B as it does so (see Fig.\ \ref{tidal} right side). A therefore slows down, and the angular momentum of the orbit of B increases. This causes B to move to a larger semi-major axis. This is actually what happens in the Earth-Moon system, where Earth's rotation period of one day is shorter than the Moon's orbital period of ca.\ 27.3~days: the Earth's rotation slows down, and the Moon moves outwards over time. This effect has a magnitude of ca.\ 38~mm increase of the Moon's semi-major axis per year, measured through laser reflectors left on the Moon during the Apollo program \citep{Chapront2002}.

\section{Activity in stellar binaries}

For binary stars in close orbits, the tidal synchronization has strong observable effects on their stellar activity. After the stellar rotation of both stars has synchronized with the orbital period of the binary, it stays locked to this period. Even though the stars still lose some angular momentum due to the stellar wind they expel, they are kept at a high rotation rate due to this tidal locking, i.e.\ over long time scales the orbital distance of the stellar binary decreases as angular momentum is lost \citep{Stepien1995}. For the activity, this means that even though the stars in a binary may have a relatively old age, they are still rotating at a period of a few days and are similarly magnetically active as a single star of that rotation period systems. 

In addition to the magnetic activity from the rotation of the individual stars, there can be interactions of the stellar magnetic fields with each other, such as magnetic loops connecting the two stars. This can lead to further magnetic activity effects \citep{Siarkowski1996, Peterson2010}.

The activity of stars in close binaries has been investigated thoroughly in various activity observables. Flare rates of M dwarfs in close pairs with white dwarfs have been found to be higher than for single M dwarfs \citep{Morgan2016}, and also their ambient activity, i.e.\ the overall activity level outside of time-resolved flares, which is thought to be a superposition of smaller-scale activity events, is high \citep{Morgan2012}. An interesting observation is that in systems with somewhat larger orbital distances of a few AU, where tidal synchronization should not be relevant, there is still an elevated activity observed \citep{Meibom2007}; this may be due to differences in the stellar formation and a different rotation period with which these moderate-distance binaries set out on the main sequence. Large-distance binaries (semi-major axes of several hundred AU) do not show this effect. But generally, activity indicators are found to be high for tidally interacting close binaries with periods of a few days \citep{Schrijver1991}.

\section{Activity in planet-hosting stars}

Thinking of a star-planet system as a scaled-down version of stellar binaries, with a very small mass ratio of the components, lets one expect that there may be relevant tidal effects as well. One active line of research in the exoplanetary field is the study of how exoplanetary orbits evolve over time, and how quickly exoplanets may spiral into their host stars \citep{Penev2011, Jackson2010}. The tidal quality factor of stars, which specifies how quickly the kinetic energy of tidal deformations and waves is dissipated, is not well constrained yet by current theories and observations, and requires further study \citep{Zahn2008, Ogilvie2007, Penev2011}. A ``smoking gun'' of an inspiraling exoplanet in the form of an actual measurement of a decreasing orbital period is yet to be found.

Concerning the tidal effects on the activity of the host star, initial theoretical studies were performed early-on. The two main interaction scenarios were identified as tidal interaction and magnetic interaction \citep{Cuntz2000}. Magnetic interaction is expected to follow scenarios either similar to loop interactions in close binaries, or similar to the Jupiter-Io unipolar inductor interaction. For tidal interaction, both the general tidal spin-up (or spin-down) of a star and activity effects due to a tidally induced increased turbulence in the outer convection layer of the star were proposed.

The observational search for star-planet interactions has been challenging. Initial detections of magnetic star-planet interaction were reported for two out of 13 stars with Hot Jupiters, where the chromospheric emission in the Ca~II lines was observed to modulate with the planetary orbital period, not the stellar rotation  period \citep{Shkolnik2005}. Later observational campaigns of those targets, however, showed that during those later epochs a modulation with the stellar rotation period was present \citep{Shkolnik2008, PoppenhaegerLenz2011}. Other magnetic effects like flare triggering or hot spots in the stellar chromosphere and corona were expected from theoretical investigations \citep{Lanza2008, Cohen2009}. For the Hot Jupiter host HD~189733, several small flares in X-rays and the UV were observed during the time shortly after the secondary transit \citep{Pillitteri2011}. As the orbit of that planet is circular and not eccentric, it is not obvious why a certain phase of the orbit should show preferential flaring (as opposed to consistent flaring during one full half of the orbit when a stellar hot spot would be visible). Later observations and modelling suggested that a plasma trail of infalling material from the planet onto the star may be the source of the high-energy emission, with the largest viewing cross-section shortly after the secondary transit \citep{Pillitteri2014}. Another possible magnetic interaction effects has been reported for the system HD~17156, which hosts a Jupiter in a strongly eccentric orbit. The system showed elevated X-ray emission during two periastron passages of the planet, and low X-ray emission during two apoastron passages \citep{Maggio2015}. This may be analogous to colliding magnetospheres observed for some young binary stars in eccentric orbits \citep{Getman2011, Getman2016}.

Systematic investigations of stellar activity in larger samples of planet-hosting stars have been performed. While initial studies observed a trend of stars with close-in and massive planets to be more active than stars with small and far-away planets \citep{Kashyap2008}, these trends have simultaneously a large scatter over the whole sample \citep{Poppenhaeger2010} and can in part be traced back to selection effects from the efficiency of planet-detection methods for active and inactive stars \citep{Poppenhaeger2011}. Some effects, especially for extremely close and massive planets, seem to be still present when strictly controlling for the spectral type of the sample stars \citep{Miller2015}; however, not all Hot Jupiters necessarily have an active host star \citep{Poppenhaeger2009, Miller2012, Pillitteri2014WASP18}. 

\begin{figure}
\includegraphics[width=\textwidth]{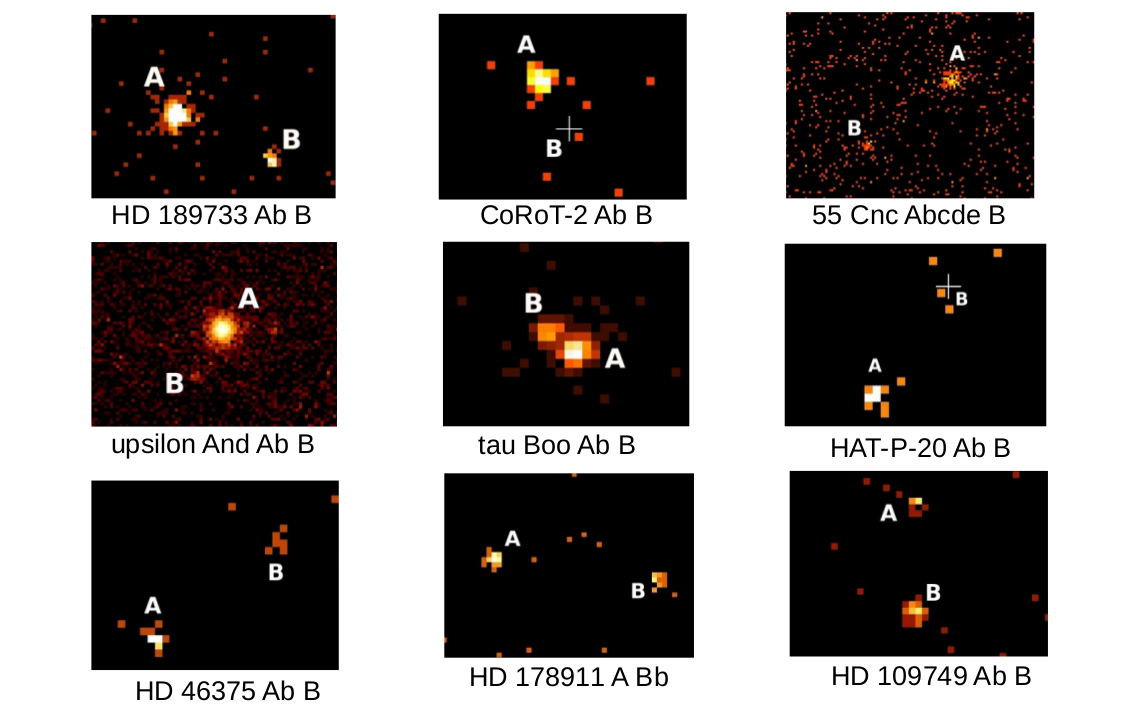}
\caption{Examples of several wide binary systems where one star hosts a known planet, observed in the X-ray band with \textit{Chandra} and \textit{XMM-Newton}.}
\label{spi}
\end{figure}

One important point is that even if planets may increase the stellar activity through some form of star-planet interaction, the magnetic braking of the star due to its stellar wind is going on at the same time \citep{Penev2012}. Taking into acocunt the age of a star-planet system is therefore crucial in order to distinguish whether a star is active because it is relatively young (and would therefore be active no matter if there was a planet or not), or if the star is actually old and is only active because it has been influenced by its planet. Unfortunately, ages for single field stars with ages over a gigayear are hard to estimate \citep{Soderblom2010review}. One way around this problem is using wide stellar binaries in which one of the stars hosts a known planet. In a wide stellar binary, the two stars will have the same age, and their activity levels should be similar (after adjusting for differences due to stellar mass). If the planet-hosting star has a much higher activity level than the companion star, one can deduce that the high activity level is not due to youngness of the system, but due to a planetary influence. In a sample of 18 such systems (see some examples in Fig.\ \ref{spi}), the stars for which a strong tidal influence is expected from their planet preferentially display higher activity levels than their companion stars (\citet{Poppenhaeger2014}, Poppenhaeger et al.\ submitted). This effect is absent for stars with planets that are not expected to have a strong tidal influence on their host stars. 

Systematic effects on stellar activity can have important consequences for exoplanets: since the atmospheric mass loss of exoplanets is thought to be driven by X-ray and extreme UV irradiation \citep{Lecavelier2004, Sanz-Forcada2010}, an elevated stellar activity level can lead to higher evaporation rates for planets. Indeed, extended planetary atmospheres and/or active atmospheric escape have been observed for several exoplanets \citep{Vidal-Madjar2003, Lecavelier2010, Poppenhaeger2013, Bourrier2013, Kulow2014, Ehrenreich2015}. For small exoplanets, such evaporation may lead to the total loss of their atmosphere \citep{Lopez2013, Poppenhaeger2012}. Especially for habitability considerations, such as for planets in the habitable zones around M dwarfs, this is an important concern \citep{Segura2010}. M dwarfs can produce frequent flares even at older ages \citep{Guedel2004, RobradePoppenhaeger2010, Davenport2016}. From a stellar perspective this is particularly interesting in the fully convective M dwarf regime, where a different dynamo than in the solar case needs to be present due to the lack of a stellar radiative core, and different models have been developed to investigate the possible magnetic field structures for these stars \citep{Browning2008, Yadav2015}. Especially since a habitable-zone exoplanet has been detected for the nearest neighbor of the Sun \citep{Anglada-Escude2016}, the fully convective M dwarf Proxima Centauri, investigations of stellar activity and its impact on exoplanet habitability will continue to be a prime concern for studying near-by exoplanets.

\section{Conclusion}

Magnetic activity is not only an interesting stellar phenomenon, but also an important topic for exoplanets. Tidal influences on stellar activity are well-known in stellar binaries, and there is some observational evidence accumulating that also massive planets in close-in orbits can influence the stellar activity. Further investigations into the observational magnitudes of tidal effects as well as into stellar tidal quality factors will be necessary for understanding of these systems.


\end{document}